\documentclass[twocolumn,showpacs,amsmath,amssymb,PRL]{revtex4}
\usepackage{graphicx}% Include figure files
\usepackage{dcolumn}% Align table columns on decimal point
\usepackage{bm}% bold math

\begin{document}

%\preprint{ELDOR 9}

\title{Spin-Dependent Recombination between Phosphorus Donors in Silicon and Si/SiO$_2$ Interface States Investigated with Pulsed Electrically Detected Electron Double Resonance}% Force line breaks with \\

\author{Felix Hoehne}
\email[corresponding author, email: ]{hoehne@wsi.tum.de} \affiliation{Walter
Schottky Institut, Technische Universit\"{a}t M\"{u}nchen, Am Coulombwall 3,
85748 Garching, Germany}
\author{Hans Huebl}
\altaffiliation{present address: Walther-Meissner-Institut, Bayerische Akademie der Wissenschaften, Walther-Meissner-Strasse 8, 85748 Garching, Germany}
\affiliation{Walter Schottky Institut, Technische Universit\"{a}t
M\"{u}nchen, Am Coulombwall 3, 85748 Garching, Germany}
\author{Bastian Galler}
\affiliation{Walter Schottky Institut, Technische Universit\"{a}t
M\"{u}nchen, Am Coulombwall 3, 85748 Garching, Germany}
\author{Martin Stutzmann}
\affiliation{Walter Schottky Institut, Technische Universit\"{a}t
M\"{u}nchen, Am Coulombwall 3, 85748 Garching, Germany}

\author{Martin S.~Brandt}
\affiliation{Walter Schottky Institut, Technische Universit\"{a}t
M\"{u}nchen, Am Coulombwall 3, 85748 Garching, Germany}

\date{\today}% It is always \today, today,

\begin{abstract}
We investigate the spin species relevant for the spin-dependent
recombination used for the electrical readout of coherent spin manipulation in phosphorus-doped silicon. Via a
multi-frequency pump-probe experiment in pulsed electrically
detected magnetic resonance, we demonstrate that the dominant
spin-dependent recombination transition occurs between phosphorus
donors and Si/SiO$_2$ interface states. Combining pulses at different
microwave frequencies allows us to selectively address the two spin subsystems
participating in the recombination process and to coherently manipulate and detect the relative spin orientation of the two recombination partners.
\end{abstract}

% 71.55.Cn = Elemental Semiconductors
% 03.67.Lx = Quantum Computation
% 76.30.-v = Electron Paramagnetic Resonance and relaxation
% 83.85.St = stress relaxation
%73.50.Gr Charge carriers: generation, recombination, lifetime, trapping, mean free paths (thin films)

%Use showkeys class option if keyword

\maketitle

% ========================= INTRODUCTION ============================================
Electron spin resonance (ESR) is a well known tool to manipulate
electron spins in semiconductors. Due to the limited detection
sensitivity of ESR \cite{Maier97}, typically electrical
\cite{Koppens06, Mccamey06} and optical detection schemes of spin resonance
\cite{Koehler93, Wrachtrup93} are favored to detect small numbers of
spins. In both approaches, the spin state is transferred to a
charge or photon state, respectively, and in electrical~\cite{Boehme03II} as well
as in optical detection~\cite{Childress06}, the coherent
manipulation of spin states can be monitored. In spin-dependent photoconductivity, the
spin-to-charge transfer is typically achieved via a spin-dependent process
governed by the Pauli principle involving two paramagnetic
states. While we can in principle distinguish between weakly and strongly coupled spin pairs via the Rabi frequencies~\cite{Herring09}, the identification of correlated states has only been achieved indirectly in electrically detected magnetic
resonance (EDMR) until now~\cite{Dersch83, Huebl08}. In this
paper, we demonstrate that pulsed EDMR can be used to directly identify the partners participating in a
recombination process. This is achieved by individually addressing
the different partners during the EDMR pulse sequence via
irradiation with microwaves at different frequencies.

\begin{figure}
\includegraphics[height=7.0cm]{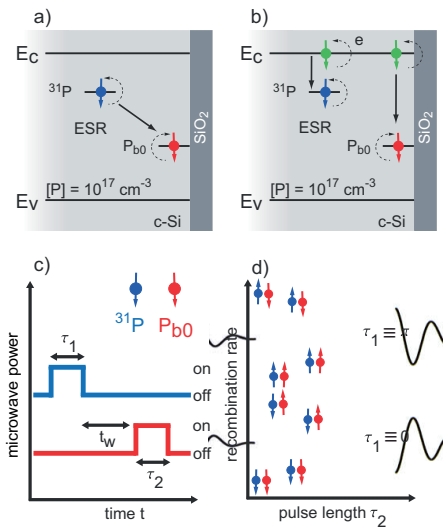}
\caption{\label{fig:DidaktikSkizzeV5} (color online) a) Spin-dependent recombination step from the $^{31}$P donor to the
$\rm{P_{b0}}$ center. In contrast, the recombination in b) involves conduction band
electrons which spin-dependently recombine with $^{31}$P and $\rm{P_{b0}}$ not involving a direct
transition between $^{31}$P and $\rm{P_{b0}}$. c) Pulse scheme used in the
multi-frequency pump-probe experiment. For the direct recombination involving $^{31}$P and $\rm{P_{b0}}$, panel d) depicts the expected Rabi oscillations in the
recombination rate induced by a probe pulse with length $\tau_2$ on the $\rm{P_{b0}}$ spin system for two
different starting conditions when no pump pulse and when a $\pi$-pump pulse has been applied to the $^{31}$P spin system.}
\end{figure}

We have performed this proof-of-principle experiment on an example
of current interest for EDMR, the read-out of phosphorus donor spin states in Si~\cite{Stegner06}.
Prominent signatures of a continuous wave (cw) experiment on heterostructures of phosphorus-doped Si and SiO$_2$ used for such a read-out can be attributed to $^{31}$P donor spins and dangling bond states $\rm{P_{b0}}$ at the Si/SiO$_2$ interface~\cite{Stesmans98}. Sometimes, a weak feature at $g~=~1.999(1)$, compatible with the spin resonance of conduction band electrons or exchange coupled $^{31}$P donors, is observed as well~\cite{Young97, Feher55}. A particular model for the spin-dependent
process monitored in these experiments is the transition from the $^{31}$P donor to the
$\rm{P_{b0}}$ state as sketched in Fig.~\ref{fig:DidaktikSkizzeV5}~a)~\cite{Stegner06}. An alternative process is depicted in Fig.~\ref{fig:DidaktikSkizzeV5}~b), showing the parallel spin-dependent transition from conduction band electrons, denoted e, to the $^{31}$P donors and the $\rm{P_{b0}}$ centers, which would result in a similar cw EDMR signature of Shockley-Read-Hall recombination~\cite{Shockley52}.
However, other mechanisms that could give rise to the observed cw resonances can also be envisaged such as scattering of conduction electrons at neutral $^{31}$P donors~\cite{Ghosh92}, capture and emission of conduction band electrons by neutral $^{31}$P donors~\cite{Thornton73,McCamey08I,Morley08}, donor-acceptor pair recombination~\cite{Stich95} and tunneling between $\rm{P_{b0}}$ states~\cite{McCamey08I}. Using cw EDMR experiments only, an identification of the process is complicated and if at all can only be achieved e.g.~by studying the dependence of the EDMR signal on magnetic field, temperature or dynamical parameters such as the phase shift observed in lock-in detection~\cite{Dersch83}. 
Here, we employ
multi-frequency pulsed EDMR, similar to electron double resonance
(ELDOR) techniques in conventional ESR, to perform pump-probe experiments identifying the dominant recombination partners.

Before describing the experimental results we will discuss the measurement scheme used in the experiments. 
As a result of the spin-dependent recombination processes sketched in Fig.~\ref{fig:DidaktikSkizzeV5}, doubly occupied diamagnetic states are formed,
which are either $^{31}$P$^-$ or $\rm{P_{b0}}^-$. In both cases, the Pauli
principle demands that these states are in a spin singlet. Therefore,
initial pairs of recombination partners (be it $^{31}$P-$\rm{P_{b0}}$ or e-$^{31}$P and
e-$\rm{P_{b0}}$ pairs) that are in an antiparallel spin state will recombine faster than
parallel spin pairs, leading to an occupation of recombination partners with parallel spins higher than
in equilibrium. In a pulsed single frequency magnetic
resonance excitation scheme, one of the spin species is rotated
selectively while the other remains unaltered if the two spins are weakly coupled. Therefore, parallel
spin states are transformed to antiparallel states depending on the pulse
length of the excitation pulse. As
shown in the lower part of Fig.~\ref{fig:DidaktikSkizzeV5}~d) for microwave pulses resonant with the $\rm{P_{b0}}$ spins, the corresponding
recombination rate will oscillate as a function of the microwave
pulse length $\tau_2$ leading to Rabi oscillations. In particular at $\tau_2~=~0$, the recombination rate will
be low due to the dominant parallel
configuration. However, if a preceding microwave pulse selectively
rotates the partner in the spin pair, the initial parallel/antiparallel configuration is changed, which is then reflected
by a change of amplitude or even an inversion of the corresponding Rabi oscillations. As an example, for
the case of a direct $^{31}$P-$\rm{P_{b0}}$ recombination process the upper part of
Fig.~\ref{fig:DidaktikSkizzeV5}~d) shows the expected Rabi
oscillations measured on the $\rm{P_{b0}}$ spins after inverting the $^{31}$P spins with a $\pi$ pulse, which are inverted when compared to the Rabi oscillations when no initial pump pulse was applied to the $^{31}$P system.
In contrast, if the recombination path is as indicated in
Fig.~\ref{fig:DidaktikSkizzeV5}~b), where the recombination is
directly from the conduction band to either the $^{31}$P or $\rm{P_{b0}}$, a
preceding pulse on the spin species not involved
in the spin-dependent recombination step (e.g. a pulse on the $^{31}$P when measuring the Rabi oscillations on the $\rm{P_{b0}}$) should not change the
initial parallel/antiparallel ratio of the spin pair giving rise to the spin selection (the e-$\rm{P_{b0}}$ pair in this example). Therefore, the Rabi oscillations should remain unchanged in this case.

% ----------------------FIGURE -----------------------------------
% ----------------------FIGURE -----------------------------------
% 

\begin{figure}
\includegraphics[width=8.5cm]{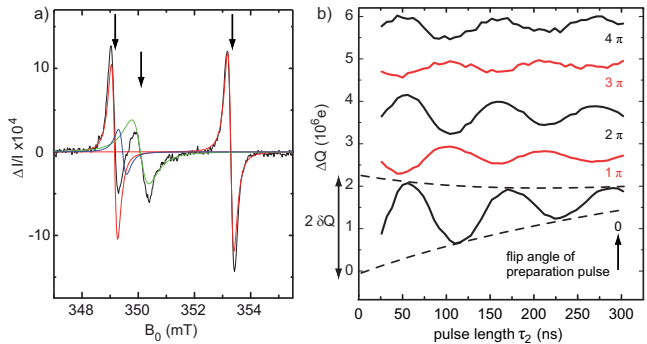}
\caption{\label{fig:GraphRabisIneV4} (color online) a) Swept-field cw EDMR spectrum of the relative current change $\Delta I/I$ (black line) at 6~K showing the two $^{31}$P hyperfine lines (red) and the two P$_{\rm{b0}}$ lines (blue and green). The colored lines are fits using a derivative Lorentzian line shape. In b), Rabi
oscillations of the P$_{\rm{b0}}$ spins for several flipping angles of
the first pulse on both $^{31}$P resonances in multiples of $\pi$ are depicted. The
Rabi oscillations for $\tau_{1}~\equiv~\pi$ and $\tau_{1}~\equiv~3\pi$ are
inverted (solid red lines) as expected for a spin-dependent
recombination involving these two spin species.}
\end{figure}

The sample investigated is a 15~nm thick silicon layer with (001) surface doped with
phosphorus at [P]=$10^{17}$~cm$^{-3}$ grown on top of a 500~nm thick
nominally intrinsic buffer deposited by chemical vapor deposition
on a Si:B wafer (30~$\Omega$cm)~\cite{huebl06}. The EDMR measurements were performed for a magnetic field
of $B_{0}$$\left|\right|$[110] in a Bruker ESR
dielectric microwave resonator with a quality factor of $Q~\approx$~100 at 6~K under illumination with the white light of a
tungsten lamp. In the cw EDMR spectrum shown in Fig.~\ref{fig:GraphRabisIneV4}~a) we observe two $^{31}$P donor lines with $g$~=~1.9985 and a hyperfine (hf) splitting of $\approx$~4.2~mT~\cite{Feher55} and two $\rm{P_{b0}}$ resonances at $g$~=~2.005(1) and 2.009(1)~\cite{Stesmans98}. The pulsed experiments were performed at a constant $B_0$~$\approx$~349.1~mT using three different microwave frequencies to excite magnetic resonance, one for the $g~=~$2.005 P$_{\rm{b0}}$ resonance ($f_{\rm{P_{b0}}}~=~$9.7938~GHz) spectrally better resolved from the low-field $^{31}$P resonance and two for the two $^{31}$P
hyperfine-split lines ($f_{P_h}~=~9.70508$~GHz,
$f_{P_l}~=~9.8202$~GHz). The corresponding spectral positions in the cw EDMR spectrum are marked in Fig.~\ref{fig:GraphRabisIneV4}~a). The two
microwave frequencies for the hyperfine-split $^{31}$P resonances
are adjusted in intensity to obtain matching Rabi frequencies corresponding to a
$\pi$ pulse time of $\approx$~37~ns, longer than the shortest possible pulse length of 20~ns in our setup. For electrical access,
interdigit Cr/Au contacts with 20~$\mu$m contact distance define
the active area of the device with $2\times$2.25~mm$^{2}$,
corresponding to approximately $10^{10}$ P spins. During the
experiment the metal-semiconductor-metal structure is biased in the ohmic region with 22~mV, resulting in a current of
$\approx$~50~$\mu$A. The amplified current transient induced by the microwave pulses is high-pass filtered ($f_{3dB}$=30~kHz) and recorded with a digital storage oscilloscope. To obtain a sufficient
signal-to-noise ratio, the experiment is repeated with
a repetition time of $250~\mu$s which is large compared to typical relaxation times~\cite{Huebl08, Paik09}. The recorded current transients are
box-car integrated from $2-11~\mu$s, giving an integrated charge
$\Delta Q$ which is directly proportional to the recombination rate at the end of the
microwave pulse sequence \cite{Boehme03,Stegner06}.

%
% ----------------------FIGURE -----------------------------------
% ----------------------FIGURE -----------------------------------
%
We now apply the pump-probe sequence sketched in
Fig.~\ref{fig:DidaktikSkizzeV5}~c). The first pulse with length
$\tau_{1}$ addresses both hyperfine-split $^{31}$P ensembles. After
$t_{w}$=30~ns, the second pulse with length $\tau_{2}$ is applied at
the P$_{\rm{b0}}$ resonance frequency. In
Fig.~\ref{fig:GraphRabisIneV4} the integrated current transient is
plotted as a function of $\tau_{2}$ for different pulse lengths
$\tau_{1}$ of the preparation pulse. For $\tau_{1}$=0~ns we observe
Rabi oscillations on the P$_{\rm{b0}}$ center as expected from
previous experiments~\cite{Stegner06}. The decay time constant of 250~ns can be attributed mainly to the inhomogeneity of the microwave $B_1$ field in our resonator. When we change $\tau_{1}$ to
37~ns, which corresponds to a $\pi$ pulse on the P donor spins, we
see an inversion of the Rabi oscillations on the P$_{\rm{b0}}$. A
further increase of $\tau_{1}$ to 76~ns ($\approx$~2$\pi$) again
inverts the Rabi oscillations, resembling the situation for
$\tau_{1}$=0~ns. This oscillatory behavior continues for
$\tau_{1}$ times up to 146~ns ($\approx$~4$\pi$). 

\begin{figure}
\includegraphics[width=8.5cm]{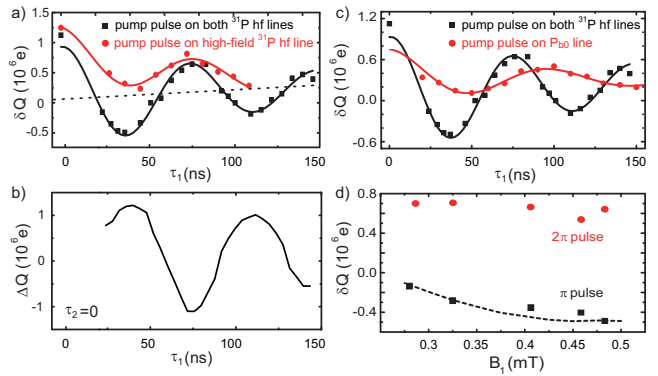}
\caption{\label{fig:RabiAmpspaper} (color online) Panel a) shows
the amplitude $\delta Q$ of the Rabi oscillations on the P$_{\rm{b0}}$ spins
vs. the pulse length $\tau_{1}$ of the pump pulse on both $^{31}$P resonances (black squares) and on the high-field $^{31}$P resonance only (red dots). The solid lines are fits with an exponentially damped cosine
plus a linear background (black dotted line for excitation of both $^{31}$P resonances). b) Rabi
oscillations excited on both $^{31}$P hyperfine lines without subsequent probe pulse. c) Comparison of $\delta Q$ for pumping on both $^{31}$P lines (black) and on the P$_{\rm{b0}}$ resonance (red). d) Amplitude $\delta Q$ of the Rabi oscillations on the P$_{\rm{b0}}$ spins 
as a function of the microwave $B_1$ field amplitude of $\pi$ and $2\pi$ pump pulses on both $^{31}$P hyperfine lines. The dashed line is a numerical simulation taking power broadening effects into account.}
\end{figure}

Figure~\ref{fig:RabiAmpspaper}~a) shows the
amplitude of the Rabi oscillations $\delta Q$ as defined in Fig.~\ref{fig:GraphRabisIneV4} b) on the P$_{\rm{b0}}$ center as a
function of the pulse length $\tau_{1}$ (black solid squares). To
extract the Rabi amplitude $\delta Q$ from the data in
Fig.~\ref{fig:GraphRabisIneV4}, each oscillation is fitted by an
exponentially damped cosine plus a linear background.
The Rabi amplitudes $\delta Q$ oscillate with a period of
$75.7$$\pm$1.3~ns and decrease exponentially with increasing
pulse length $\tau_{1}$. This oscillation period is in good agreement with the single frequency Rabi oscillations $\Delta Q$ excited on the two hyperfine-split $^{31}$P resonances without a subsequent pulse on the P$_{\rm{b0}}$ spins ($\tau_2$~=~0) 
shown in Fig.~\ref{fig:RabiAmpspaper}~b) with an oscillation period
of $73.7$$\pm$0.5~ns.

The clear inversion of the Rabi oscillations measured on P$\rm{_{b0}}$ indicates that the formation and recombination of pairs involving $^{31}$P and
P$\rm{_{b0}}$ (Fig.~\ref{fig:DidaktikSkizzeV5}~a)) is the spin-dependent recombination process observed in EDMR under the magnetic field and temperature conditions used here, in contrast to Ref.~\cite{McCamey08I} which studies EDMR at 8~T. The quantitative analysis performed below allows us to conclude that within an error margin of $\approx$~10~$\%$ the EDMR signal amplitude observed is completely caused by this process. 

%
% ----------------------FIGURE -----------------------------------
% ----------------------FIGURE -----------------------------------
%
The inverted Rabi oscillations for $\tau_{1}~\equiv~\pi$ and $\tau_{1}~\equiv~3\pi$ in Fig.~\ref{fig:GraphRabisIneV4} have a smaller amplitude compared to those for $\tau_{1}~\equiv~2\pi$ and $\tau_{1}~\equiv~4\pi$ in contrast to the expected monotonous decay with longer pulse length $\tau_1$. This can also be seen in Fig.~\ref{fig:RabiAmpspaper}~a) as the constant contribution to the linear offset (dotted black line). 
There are two effects causing this incomplete inversion of the Rabi oscillations. First, the bandwidth of the microwave pulses is not sufficient to excite all spins of the inhomogeneously broadened $^{31}$P lines in our experimental setup. Therefore, after applying a $\pi$ pulse most but not all of the spins of the $^{31}$P
spin ensemble are turned by $\pi$ as required for a full inversion of the Rabi
oscillations on the P$_{\rm{b0}}$ spins. Since a small fraction of the
$^{31}$P spin ensemble is not addressed by the pulse, the Rabi oscillations
of a small part of the P$_{\rm{b0}}$ spin ensemble also does not change its sign. This can be demonstrated more clearly by limiting the pump pulses to irradiation with $f_{P_h}$ only, thereby addressing half of the $^{31}$P system. As can be seen Fig.~\ref{fig:RabiAmpspaper}~a), the variation of $\delta Q$ is now indeed about half of the variation when both $^{31}$P resonances are excited.   %This results in the offset in Fig.~\ref{fig:RabiAmpspaper} a).

Second, the P$_{\rm{b0}}$ resonances are close to the low-field $^{31}$P line (see Fig.~\ref{fig:GraphRabisIneV4} a)), so that a pump pulse on both $^{31}$P hf lines partially also excites the P$_{\rm{b0}}$ spins. This also results in an incomplete inversion of the Rabi oscillations after a $\pi$ pump pulse. To estimate this offset quantitatively, we describe the linewidths of the $^{31}$P resonances and the two P$_{\rm{b0}}$ resonances, which are in fact caused mostly by superhyperfine interactions with $^{29}$Si nuclei, by corresponding Gaussian distributions of $g$-factors with full width at half maximum
of $\Delta g_{\rm{P}}$=0.001, $\Delta g_{\rm{Pb0}}$=0.0016 and $\Delta g'_{\rm{Pb0}}$=0.0008. From this, we
estimate the fraction of $^{31}$P spins turned by a $\pi$ pulse of 37~ns (corresponding to a microwave magnetic field $B_{1}$~=~0.48~mT) to $\approx$~0.9 and the fraction of P$_{\rm{b0}}$ spins to $\approx$~0.1 which accounts for the constant offset in Fig.~\ref{fig:RabiAmpspaper}~a) quantitatively. This is corrobrated by the pump experiments on the spectrally better resolved high-field $^{31}$P resonance only, where an amplitude of the oscillations of $\delta Q$ of 4.9~$\times~10^5$~e is obtained from the fit in Fig.~\ref{fig:RabiAmpspaper}~a), while 8.2~$\times~10^5$~e is found when both $^{31}$P resonances are excited. Comparison of these two values allows to determine the fraction of P$_{\rm{b0}}$ spins turned by exciting the less resolved $^{31}$P resonance to $\approx$~0.2 in reasonable agreement with the value estimated earlier. This discussion shows that higher magnetic fields $B_0$ removing the spectral overlap of resonances would be beneficial. 

Furthermore, the model is supported by repeating the experiment at lower powers of the microwave pulses on the two $^{31}$P ensembles as shown in Fig.~\ref{fig:RabiAmpspaper}~d). The amplitude $\delta Q$ after a $\pi$ pump pulse decreases with decreasing $B_1$ field whereas it remains almost constant for a $2\pi$ pump pulse as expected. The lower $B_1$ fields
lower the fraction of $^{31}$P spins affected by the pulses which reduces the inversion. The dashed line is a numerical simulation taking these power broadening effects into account. The origin of the additional slight increase of the background in
Fig.~\ref{fig:RabiAmpspaper}~a) can similarly at least qualitatively be
explained taking into account that the spectral width of the microwave pulses decreases with increasing pulse length also when keeping the microwave power constant. Therefore, the fraction of $^{31}$P spins
addressed by the pulses is reduced when $\tau_1$ is increased resulting in an increasing
background.

If the $^{31}$P-$\rm{P_{b0}}$ spin pair recombination indeed takes place, an exchange of the $^{31}$P and $\rm{P_{b0}}$ pulses in the pulse sequence should result in the same pulsed EDMR signature. We performed
this experiment by applying the pump pulse to the $\rm{P_{b0}}$ spin species and monitoring Rabi oscillations on the high-field $^{31}$P resonance. As shown in Fig.~\ref{fig:RabiAmpspaper}~c), the amplitude $\delta Q$ of the Rabi oscillations on $^{31}$P oscillates with a period of $\approx$~100~ns characteristic for the length of $2\pi$ pulses on the $\rm{P_{b0}}$ system (c.f. Fig.~\ref{fig:GraphRabisIneV4}~b)). However, also in this case no inversion of the Rabi oscillations after a $\pi$ pulse was observed. The $g$-factor
distribution of the two P$_{\rm{b0}}$ lines is wider compared to the $^{31}$P spins and therefore only a smaller fraction of $\approx$~0.5 of all $\rm{P_{b0}}$ spins is
addressed by the first microwave pulse preventing an inversion of the Rabi oscillations. 

Two models to account for the relative current change $\Delta I/I$
detected in cw EDMR experiments are usually discussed, the Lepine
model \cite{Lepine72} based on the polarisation of the spin species
participating and the Kaplan-Solomon-Mott model \cite{Kaplan78}
assuming weakly coupled pairs. While the experiments reported here
clearly demonstrate that the spin-dependent recombination step
we monitor in the Si:P epilayers takes place between the $^{31}$P and the
$\rm{P_{b0}}$ center, we cannot make conclusions on the coupling
from these experiments. Most likely, exchange interaction caused by an overlap of the two wavefunctions leads to this coupling. However, also an indirect coupling e.g.~by itinerant charge carriers such as electrons or holes cannot be excluded at this point. However, together with the recently
demonstrated capability to refocus the spin system and observe
recombination echoes electrically \cite{Huebl08} the multi-frequency pulsed EDMR reported here will allow e.g.~to
measure the $^{31}$P-$\rm{P_{b0}}$ spin coupling via the Double
Electron Electron Resonance pulse scheme. A sample with a sufficiently narrow distribution of $^{31}$P-$\rm{P_{b0}}$ coupling strengths thus
determined might allow to demonstrate the elemental two qubit
operation, the CNOT, between the electron spins of the phosphorus donor and
the $\rm{P_{b0}}$ interface state.

To summarize, we have used pulsed electrically detected electron
double resonance to investigate the spin-dependent
recombination in phosphorus doped crystalline silicon. In the
pump-probe experiment performed, we show that the rotation of the
$^{31}$P spins by a pump pulse results in an oscillating amplitude
of the Rabi oscillations detected on the P$_{\rm{b0}}$ center and vice versa. In
particular, using a $\pi$ pump pulse on both $^{31}$P resonances we observe an inversion of the
Rabi oscillations on the P$_{\rm{b0}}$ line. This interplay of the
two spin species clearly demonstrates that the spin-dependent
recombination proceeds between the phosphorus donor and a
P$_{\rm{b0}}$ center. The technique shown here is not limited to
this specific spin system, but can be applied in general to identify
the partners participating in spin-dependent transport processes.

The authors would like to thank A. R. Stegner for fruitful
discussions. The work was financially supported by DFG (Grant No. SFB 631, C3).

\clearpage
\end{document}